\documentclass[11pt,letterpaper,english,aps]{revtex4}
\usepackage[T1]{fontenc}
\usepackage[latin1]{inputenc}
\usepackage{verbatim}
\usepackage{subfigure}
\usepackage{amsmath}
\usepackage{graphicx}

\makeatletter

\usepackage{babel}
\makeatother
\begin{document}

\title{Dynamics of ultrathin metal films  on amorphous substrates}

\author{Christopher Favazza$^{\text{1,3}}$, Ramki Kalyanaraman$^{\text{1,3}}$%
\thanks{E-mail: ramkik@wuphys.wustl.edu%
} and Radhakrishna Sureshkumar$^{\text{2,3}}$%
\thanks{E-mail: suresh@wustl.edu%
}}

\affiliation{$^{\text{1}}$Department of Physics, Washington University in St. Louis, MO 63130}

\affiliation{$^{\text{2}}$Department of Energy, Environmental and Chemical Engineering, Washington University
in St. Louis, MO 63130}

\affiliation{$^{\text{3}}$Center for Materials Innovation, Washington University in St. Louis, MO 63130\thispagestyle{empty}}

\begin{abstract}
A mathematical model is developed to analyze the growth/decay rate of surface perturbations of
an ultrathin metal film on an amorphous substrate ($SiO_{2}$). The formulation combines the
approach of Mullins {[}J. Appl. Phys. v30, 77, 1959] for bulk substrates, in which curvature-driven
mass transport and surface deformation can occur by surface/volume diffusion and evaporation-condensation
processes, with that of Spencer et al. {[}Phys. Rev. Lett. v67, 26, 1991] to describe solid state
transport in thin films under epitaxial strain. Modifications of Mullins model to account for
thin film boundary conditions result in qualitatively different dispersion relationships especially
in the limit as \emph{$kh_{o}\ll1$} where \emph{$k$} is the wavenumber of the perturbation
and \emph{$h_{o}$} is the unperturbed film height. The model is applied to study the relative
rate of solid state mass transport as compared to that of liquid phase dewetting in a thin film
subjected to fast a thermal pulse. Specifically, we have recently shown that multiple cycles
of nanosecond ($ns$) pulsed laser melting and resolidification of ultrathin metal films on amorphous
substrates can lead to the formation of various types of spatially ordered nanostructures {[}Phys.
Rev. B, v75, 235439 (2007)]. The pattern formation has been attributed to the dewetting of the
thin film by a hydrodynamic instability. In such experiments the film is in the solid state during
a substantial fraction of each thermal cycle. However, results of a linear stability analysis
based on the aforementioned model suggest that solid state mass transport has negligible effect
on morphology changes of the surface. Further, a qualitative analysis of the effect of thermoelastic
stress, induced by the rapid temperature changes in the film-substrate bilayer, suggests that
stress relaxation does not appreciably contribute to surface deformation. Hence, surface deformation
caused by liquid phase instabilities is rapidly quenched-in during the cooling phase. This deformed
state is further evolved by subsequent laser pulses. These results have implications to developing
accurate computer simulations of thin film dewetting by energetic beams aimed at the manufacturing
of optically active nanoscale materials for applications including information processing, optical
devices and solar energy harvesting.\pagebreak
\end{abstract}
\maketitle

\section{Introduction}

Ion and photon beam irradiation are widely used to generate nanoscale surface features for a
variety of applications: optoelectronic, plasmonic, magnetic, bio- and chemical sensing, and
catalytic devices \cite{McHenry,Shipway,Xia,Raty}. Understanding interfacial transport mechanisms
under fast thermal processing or energetic beam irradiation is essential to accurately predicting
the spatio-temporal dynamics of film evolution that imparts the desired nanoscale surface features.
The contributions from various solid state transport processes that cause surface changes such
as flattening or grooving in \textit{bulk materials} were studied by Mullins \cite{Mullins1,Mullins2,Mullins3}.
In his formulation,  smoothening of bulk surfaces is driven by curvature effects that manifest
in solid state surface diffusion, volume diffusion, or evaporation-condensation mechanisms\cite{Mullins3}.
Mullins' theory has been widely accepted and used extensively in the analyses of surface processes
\cite{Espinosa,Cahill,Bartelt,Tritscher,Tmisit,Ogasawara}. However, Mullins' theoretical framework
is inadequate to describe solid state mass transport in \emph{ultrathin films} in which one has
to account for the effects of other driving forces on pattern formation, e.g. dispersion forces
and epitaxial stress. For instance, it is well-established that epitaxial stress has a significant
effect on surface diffusion. There is extensive literature on pattern formation in thin films
in the presence of epitaxial stresses \cite{Spencer,Lu,LeGoues,Jensen}. In contrast, the literature
on the evolution of ultrathin films on amorphous substrates, a scenario in which epitaxial stress
effects are negligible, is sparse. In this article, we develop a generally applicable mathematical
model for describing such processes and apply the model to study solid state transport in thin
films exposed to time periodic laser thermal stimulus. 

Recently, we have observed that thermal processing of ultrathin metal films by multiple instances
of a $9\, ns$ pulsed laser (with each pulse separated in time by $20\: ms$) results in a variety
of spatially ordered patterns \cite{favazza06c,Favazza06b,favazza06d}. One of the important
findings, relating to the processing conditions, was that in order for pattern formation to occur,
it was necessary to irradiate the films at a laser fluency at or above a critical value, which
was well-correlated with the threshold energy required to melt the films \cite{Trice06a}. Consequently,
mass transport was considered to occur primarily in the liquid phase and pattern formation was
attributed to a thin film hydrodynamic (TFH) dewetting instability, which arises in spinodally
unstable films when attractive long-range intermolecular dispersion forces amplify surface perturbations
\cite{favazza06d}. Thermal modeling of the film/substrate bilayer showed that during one laser
pulse, the film experienced a nearly uniform, but transient temperature field for a time scale
of $O(100\, ns)$, of which the time to reach the melt temperature during the heating phase was
$\sim1\, ns$, the liquid lifetime, $\tau_{L}$ was typically between $1-10\, ns$ and cooling
time was $\sim100\, ns$ \cite{Trice06a}. Therefore, approximately $90\%$ of the thermal cycle
following each laser pulse was spent in the cooling phase. This is important in the context of
Mullins' analysis, which suggests that solid state diffusion and evaporation-condensation processes
could influence surface deformation. Hence, it is necessary to analyze both liquid and solid
phase mass transport processes to develop a comprehensive understanding of the physics of interface
evolution in thin films processed under pulsed laser irradiation. 

In this work, we have suitably modified Mullins' original formulation \cite{Mullins3} to describe
the evolution of infinitesimally small surface perturbation of ultrathin solid films on amorphous
substrates. In particular, we have derived new results for volume diffusion transport and have
employed the approach adopted by Spencer et al. \cite{Spencer} originally for describing diffusion
in epitaxially stressed films, in order to account for dispersion forces that are influential
only in thin films. For pulsed laser experiments described in \cite{favazza06c,favazza06d,Trice06a},
we determined that there is negligible contribution to surface deformation from solid state material
redistribution by evaluating solid state mass transport rates via diffusion and evaporation-condensation
processes. These results suggest that film deformation occurring under each melting cycle in
the liquid phase is essentially \emph{quenched-in} during rapid resolidification. Subsequent
laser pulses cause progressively more interface deformation until the film ruptures. This mechanism
is consistent with experimental observations \cite{favazza06d,Trice06a}.

\section{Solid state mass transport \label{sec:Solid-state-mass}}

Local curvature differences provide the driving force for mass rearrangement through various
mechanisms in the solid state including surface diffusion, volume diffusion, and evaporation-condensation.
In a series of papers, Mullins outlined a theory to predict the mass transfer rates by the aforementioned
processes for surface deformation of bulk materials \cite{Mullins1,Mullins2,Mullins3}. For thin
films, additional driving forces need to be accounted for, such as the free energy contributions
from dispersion (van der Waals) forces and thermal strain. For films supported by amorphous substrates,
epitaxial strain is irrelevant. However, mismatch in thermal expansion coefficients of the film
and the substrate can result in thermoelastic strain. In this work, we first develop the theory
by incorporating dispersion forces, in the absence of thermal stresses. The effect of thermal
stresses is only discussed qualitatively in Sec. \ref{sec:Thermal-Stress}. 

Following Spencer et al. \cite{Spencer}, the evolution of the film height \emph{h} is described
by: \begin{equation}
\frac{\partial h}{\partial t}=B_{SD}\nabla^{2}(\gamma\kappa+\varepsilon)+V\label{eq:Davis}\end{equation}
where $B_{SD}=D_{S}\Omega^{2}\nu/k_{B}T$, $\gamma$ is the surface tension, $\kappa$ is the
local curvature, $D_{S}$ is the surface diffusion coefficient, $\Omega$ is the atomic volume,
$\nu$ is the number of atoms per unit area, $k_{B}$ is Boltzmann's constant, $T$ is the absolute
temperature, $\varepsilon$ is a free energy per volume term, and $V$ is an additional velocity
term such as a material deposition term \cite{Mullins1,Mullins2,Spencer}. In the limit that
the height of any surface fluctuations are small, the curvature can be approximated as $\kappa\approx\nabla^{2}h$.
It has already been established that dispersion forces play a key role in the evolution of a
thin metal film \cite{Suo}, especially upon laser-induced melting \cite{favazza06c,Favazza06b,favazza06d,Trice06a,Bischof}.
By including the effect of the disjoining pressure to the free energy, the height evolution equation
becomes: \begin{equation}
\frac{\partial h}{\partial t}=B_{SD}\nabla^{2}[\gamma\nabla^{2}h+\frac{A}{6\pi h^{3}}]+V\label{eq:dh/dtSD}\end{equation}

\noindent where $A$ is the Hamaker coefficient and the free energy per volume contribution due
to dispersion forces, $\frac{A}{6\pi h^{3}},$ is strictly valid for a flat film.

The effects of evaporation-condensation and volume diffusion processes can be included into the
height evolution equation as additional velocity terms, in eq. \ref{eq:Davis}. By relating local
pressure differences between the flat and perturbed states to the local curvature though the
Gibbs-Thompson formula, Mullins has shown that the evolution of a rough surface, due to evaporation-condensation
is given by: \begin{equation}
V_{EC}=(\frac{\partial h}{\partial t})_{EC}=B_{EC}\nabla^{2}h\label{eq:dh/dtEvap}\end{equation}
where $B_{EC}=p_{o}\gamma\Omega^{2}/(2\pi m)^{\frac{1}{2}}(k_{B}T$)$^{\frac{3}{2}}$, $p_{o}$
is the equilibrium vapor pressure and $m$ is the mass of the metal atom \cite{Mullins1,Mullins3}. 

As in the case of evaporation-condensation, volume diffusion can be considered and included in
the height evolution equation as another $V$ term. In the derivation below, \emph{x} and \emph{z}
denote the lateral and normal coordinates respectively such that \emph{z} = 0 and \emph{h} correspond
to the film-substrate interface and local film height respectively. The substrate is much thicker
than the film and hence considered as a semi-infinite domain. The decay rate due to volume diffusion
can be related to the local density, $\rho$ as $\frac{\partial h}{\partial t}=D_{V}\Omega(\frac{\partial\rho}{\partial z})_{z=h}$,
where $D_{V}$ is the coefficient of self-diffusion and \emph{z} denotes the coordinate normal
to the substrate plane \cite{Mullins3,Herring}. For a bulk surface ($-\infty<z\leq0$), assuming
quasi steady-state diffusion $(\nabla^{2}\rho(x,z)=0)$ and solving the Laplace equation with
boundary conditions $\rho(x,z)=\rho_{o}$ at $z\rightarrow-\infty$ and $\rho(x,z)=\rho_{o}+\rho_{o}\frac{\gamma\kappa}{k_{B}T}k^{2}\epsilon sin(kx)$
at $z=0$, where $\epsilon$ and $k$ are the amplitude and wavevector of the surface perturbation
 $h'$ respectively, Mullins derived $\frac{\partial h'}{\partial t}=B_{Vol}k^{3}h'$, where
$B_{Vol}=\frac{D_{V}\gamma\Omega}{k_{B}T}$. However, for a film of finite thickness the boundary
conditions need to be modified, specifically: $\rho(x,z)=\rho_{o}$ at $z=0$ and $\rho(x,z)=\rho_{o}+\rho_{o}\frac{\gamma\kappa}{k_{B}T}k^{2}\epsilon sin(kx)$
at $z=h$. Assuming quasi-steady state diffusion and $h=h_{o}+h'$ where $h'=\epsilon e^{\sigma_{S}t}sin(kx)$
the expression for volume diffusion becomes $(\frac{\epsilon}{h_{o}}\ll1)$: \begin{equation}
V_{Vol}=\frac{\partial h'}{\partial t}=-B_{Vol}k^{3}\frac{1+exp(-2kh_{o})}{1-exp(-2kh_{o})}h'\label{eq:dh/dtVol}\end{equation}

\noindent where $h_{o}$ is the average film thickness. 

Note that the quasi-steady state (QSS) assumption is only valid if $t\gg\frac{h_{0}^{2}}{D_{V}}$,
and even for sub-nanometer thick metal films, the time to reach QSS, $t\gg200\, ns$. However,
since $(\frac{\partial\rho}{\partial z})_{z=h}$ (the density gradient at the surface of the
film) will have its largest magnitude for short times and decrease to its steady-state value,
the decay rate represented by a quasi steady-state approximation is a limiting case. For systems
with a timescale $t<\frac{h^{2}}{D_{V}}$ the decay rate will be faster than that predicted above.
For macroscopic systems $(kh\gg1)$, eq. \ref{eq:dh/dtVol} reduces to the Mullins equation for
volume diffusion, $\partial h'/\partial t=-B_{Vol}k^{3}h'$ \cite{Mullins3}.  It is interesting
to note that in the opposite limit, i.e., for thin films with a relatively large perturbation
wavelength ($kh\ll1$) and after linearizing the exponential about $h_{o}$, the expression for
decay due to volume diffusion simplifies to:\begin{equation}
\frac{\partial h'}{\partial t}=\frac{-B_{Vol}k^{2}}{h_{o}}h'.\label{eq:dh/dtVolLim0}\end{equation}

\noindent This limit applicable for thin films is qualitatively different from that for a bulk
material, \textit{with the rate of decay varying as $k^{2}$} \textit{as compared to $k^{3}$
for the latter. Further, the thickness dependence, i.e., the decay rate varying as $\frac{1}{h_{o}}$,
is absent in the case of bulk materials.}

By combining eqs. \ref{eq:dh/dtSD}, \ref{eq:dh/dtEvap} and \ref{eq:dh/dtVol}, the evolution
equation for height perturbations $h'$$(\ll h_{o})$ can be expressed as:

\noindent \begin{equation}
\frac{\partial h'}{\partial t}={-B}_{SD}[\gamma\nabla^{4}h'+\frac{A}{2\pi h_{o}^{4}}\nabla^{2}h']-B_{Vol}k^{3}\frac{1+exp(-2kh_{o})}{1-exp(-2kh_{o})}h'+B_{EC}\nabla^{2}h'.\label{eq:dh/dtSolid}\end{equation}

\noindent Letting $h'=\epsilon e^{\sigma_{S}t}sin(kx)$, the growth/decay rate, $\sigma_{S}$
of surface perturbations in the solid state is given by:

\begin{equation}
\sigma_{S}=-B_{SD}[\gamma k^{4}-\frac{A}{2\pi h_{o}^{4}}k^{2}]-B_{Vol}k^{3}\frac{1+exp(-2kh_{o})}{1-exp(-2kh_{o})}-B_{EC}k^{2}\label{eq:sigmaSolidState}\end{equation}

\section{Hydrodynamics of thin liquid metal films \label{sec:Hydrodynamic-flow-in}}

In this section we summarize the results for the growth/decay rate of thin liquid films with
the purpose of comparing the mass transport rates in the liquid state to that in the solid state.
For a detailed description of liquid state mass transport and dewetting in thin films, see %
{}\cite{vrij,vrij2,deGennes,kondic,sharma86,Reiter}. When the Navier-Stokes equations, describing
mass transport in thin liquid films, are simplified via the lubrication approximation, a film-thickness
dependent relationship between the material flow rate and pressure gradient can be derived \cite{vrij,vrij2}.
The pressure is a combination of the Laplace pressure from capillarity and a disjoining pressure
that arises from long-range dispersion or van der Waals forces \cite{vrij,vrij2}. When the disjoining
pressure, which tends to amplify perturbations, exceeds the stabilizing effect of the surface
tension, the film is in an unstable state and perturbations will grow, eventually causing the
film to rupture. The thin film hydrodynamic equation describing this behavior is given by the
following: \begin{equation}
\frac{\partial h}{\partial t}=-\frac{1}{3\eta}\nabla\cdot(\gamma h^{3}\nabla\nabla^{2}h+\frac{A}{2\pi h}\nabla h)\label{eq:hydro}\end{equation}
where $\eta$ is the dynamic viscosity. Once again letting $h=h_{o}+\epsilon e^{\sigma_{L}t}sin(kx)$,
the growth/decay rate $\sigma_{L}$in the liquid phase is given by:\begin{equation}
\sigma_{L}=-\frac{h_{o}^{3}}{3\eta}(\gamma k^{4}-\frac{A}{2\pi h_{o}^{4}}k^{2})\label{eq:sigmaLiquid}\end{equation}
yielding the largest unstable wavevector, $k_{max}=\sqrt{\frac{A}{2\pi{\gamma h}_{o}^{4}}}$
and a fastest growing perturbation wavelength, $\lambda^{*}=\sqrt{\frac{16\pi^{3}\gamma}{A}}h_{o}^{2}$.

\section{evolution of surface perturbations in solid and liquid states \label{sec:Deformation-cool}}

\begin{figure}
\subfigure[]{\includegraphics[width=3in]{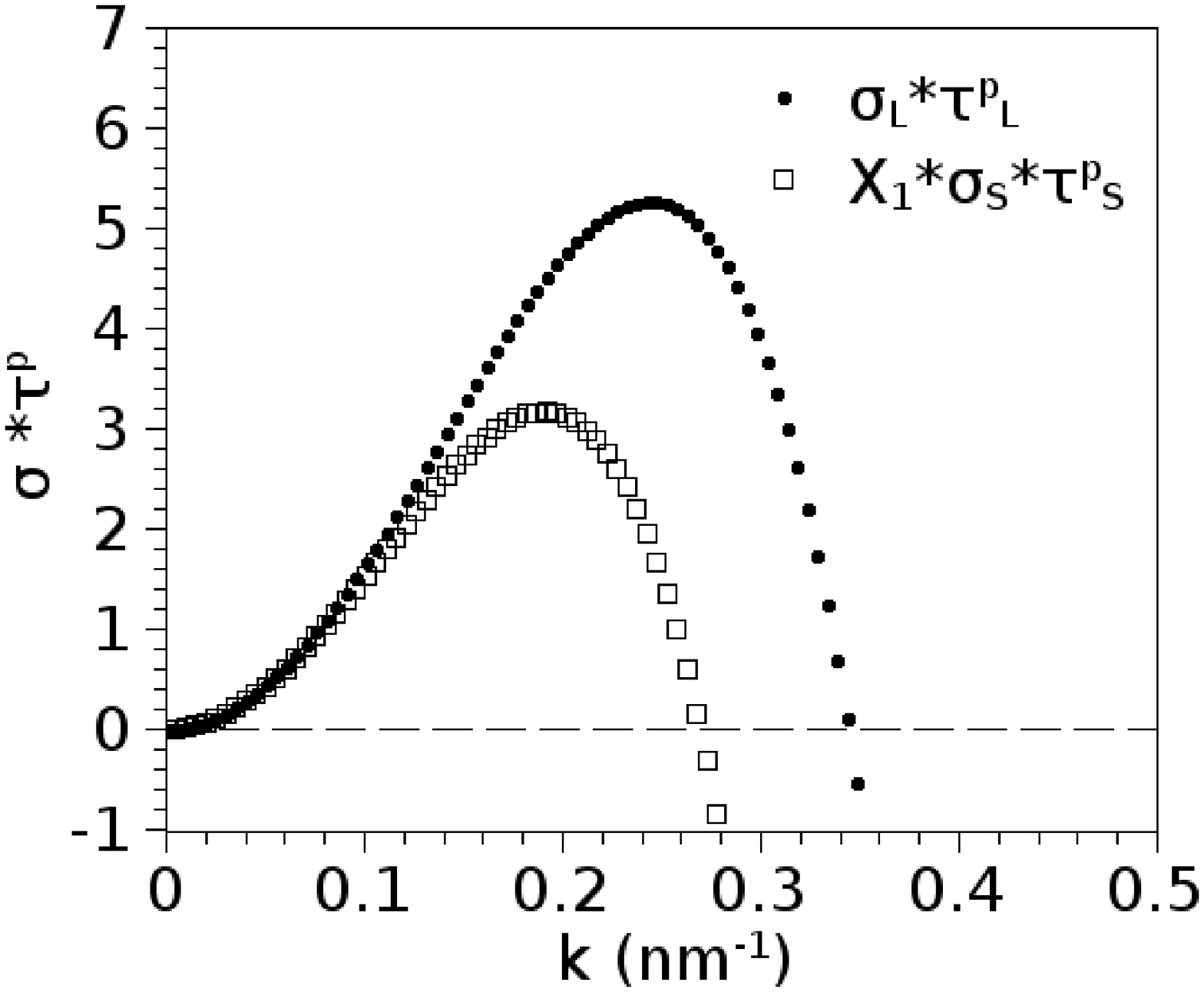}}\subfigure[]{\includegraphics[width=3in]{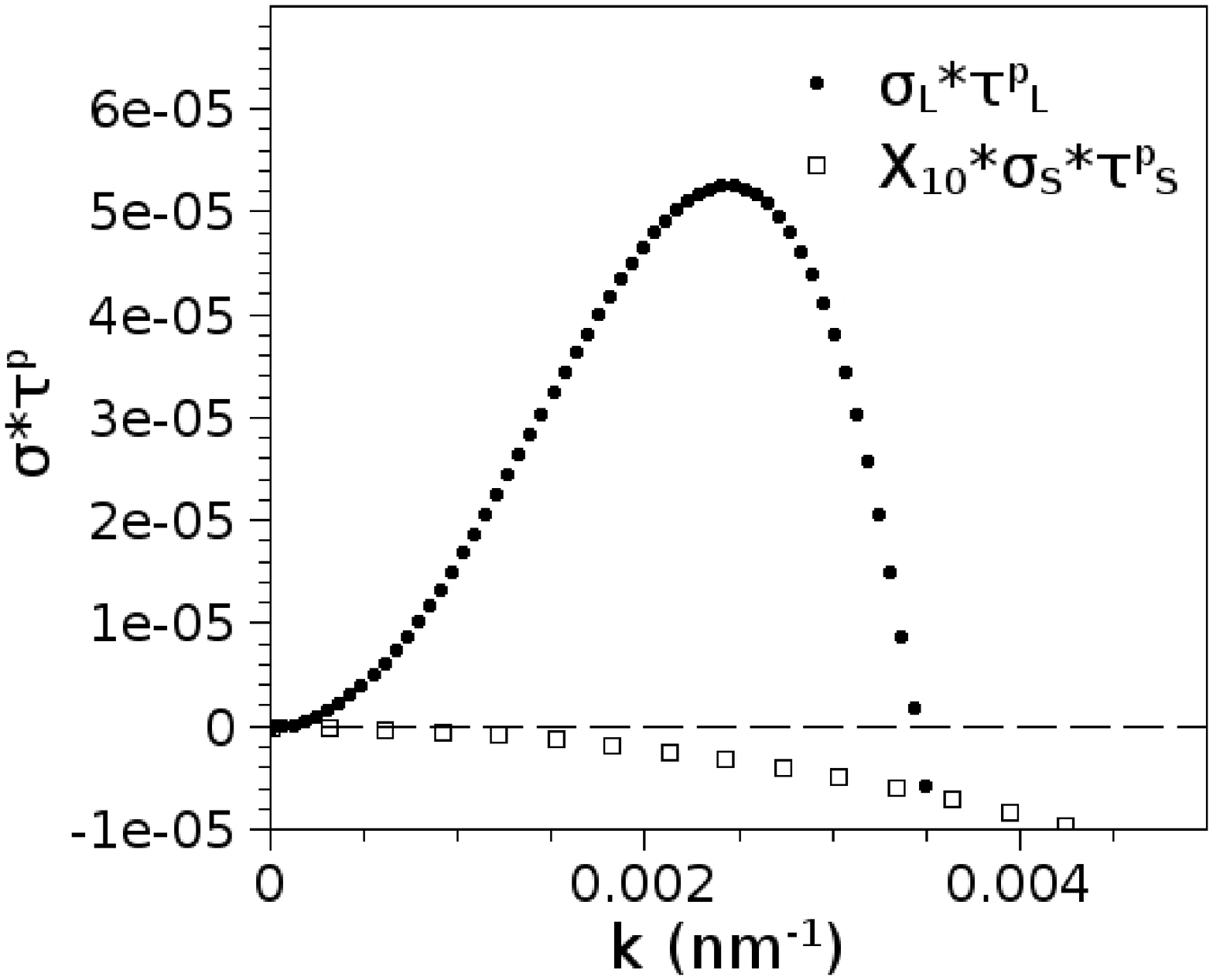}}

\caption{Plots of the dispersion relations for (a) $1\, nm$ Co film (using $A=1.4\times10^{-18}\, J$),
with the scale factor, $X_{1}=2000$ and (b) $10\, nm$ Co film where scale factor $X_{10}=100$.
In both plots $\tau_{S}^{p}=50\, ns$ and $\tau_{L}^{p}=10\, ns$. \label{fig:CoDispersionPlots}}
\end{figure}

As mentioned previously (Sec. \ref{sec:Hydrodynamic-flow-in}), analytical models and numerical
simulations have shown that for  $ns$ pulsed laser irradiation of thin metal films, the cooling
time $\tau_{cool}$ of the film is approximately $10$ times as large as the time  that the film
is in the molten state, i.e., $\tau_{cool}\sim100\, ns$ \cite{Trice06a}. At the beginning of
the cooling phase, the film is at its melting temperature. The high temperatures suggest that
solid state material transport mechanisms could be active. To estimate the relative magnitude
of the effect that solid state mass transport could have on deformation of the film during the
cooling period, we employed eqs. \ref{eq:sigmaSolidState} and \ref{eq:sigmaLiquid}. For the
values of $D_{S}$ and $D_{V}$, we used the models of Flynn for surface and volume diffusion
in which $D_{S}=5\times10^{-4}exp(-6T_{M}/T)\,\frac{cm^{2}}{s}$ and $D_{V}=0.3exp(-17T_{M}/T)\,\frac{cm^{2}}{s}$
\cite{Flynn}. In order to gain quantitative insight on the effect of mass transport in each
phase, we have compared the growth/decay rate of a perturbation multiplied by the timescale during
which it is applicable. Since diffusion and evaporation processes are disproportionally influenced
by the time spent at the peak temperature, we have used an overestimate of the time at the peak
temperature of $\tau_{S}^{p}=50\, ns$ in order to find an upper limit of the contribution from
the solid state transport mechanisms. In Fig. \ref{fig:CoDispersionPlots}(a-b) we have plotted
the magnitude of $\sigma_{S}*\tau_{S}^{p}$ and $\sigma_{L}*\tau_{L}^{p}$ as a function of perturbation
wavevector $k$ for material redistribution in both phases, where $\tau_{L}^{p}=10\, ns$ is
the typical liquid lifetime. Due to the thickness dependency of several of these mechanisms,
we have compared transport mechanisms in a $1\, nm$ and a $10\, nm$ Co film. 

\begin{figure}
\subfigure[]{\includegraphics[width=3in]{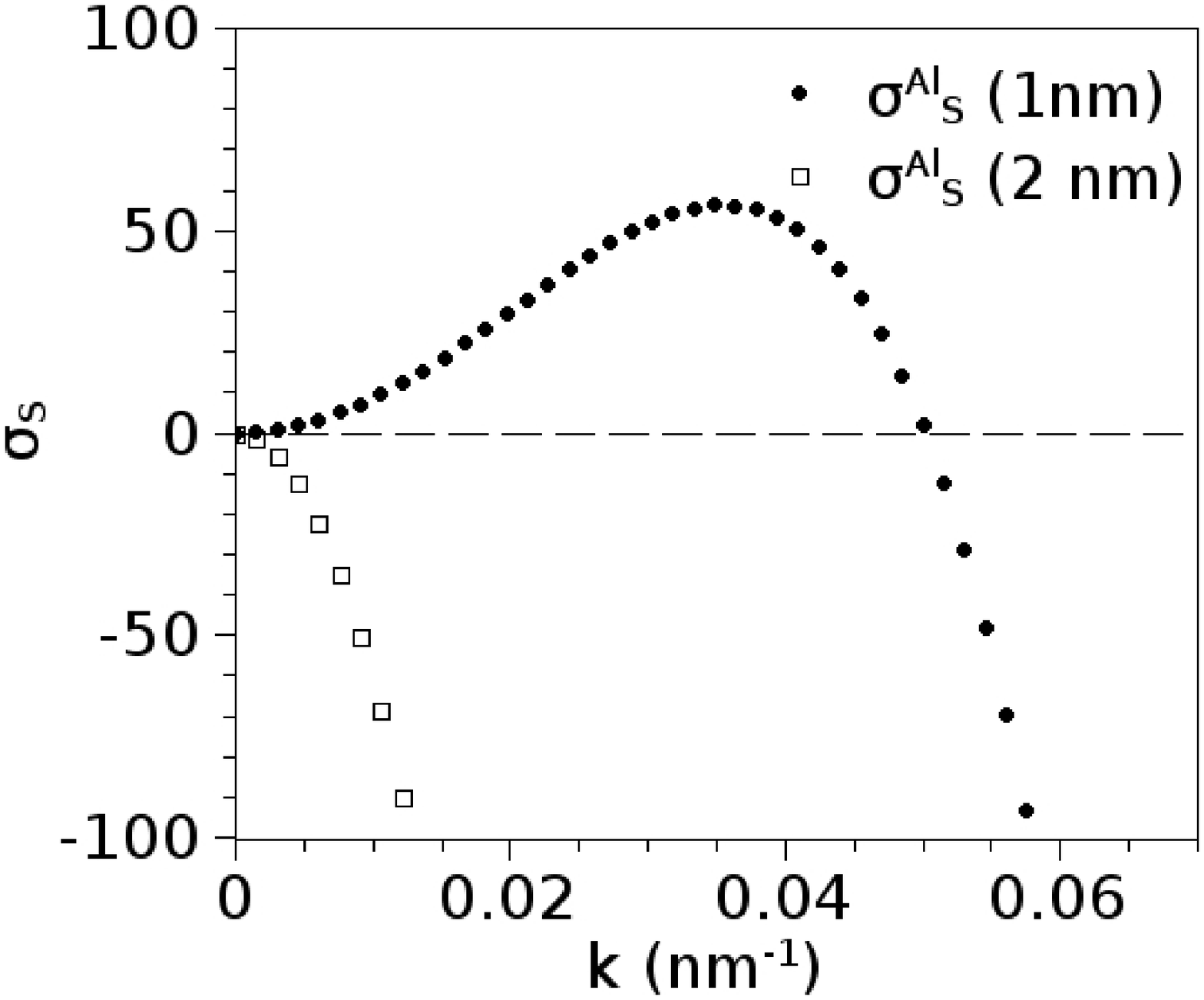}}\subfigure[]{\includegraphics[width=3in]{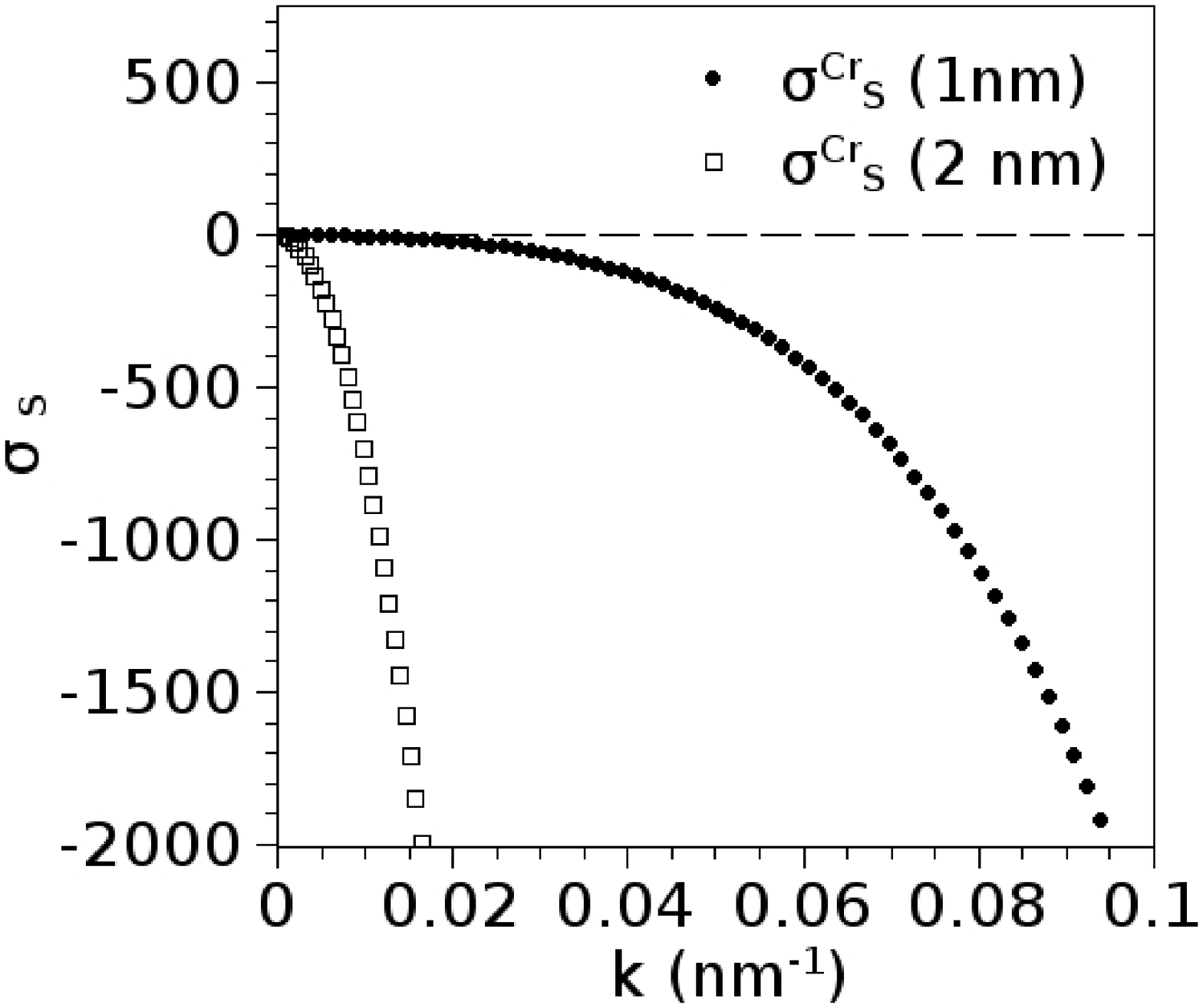}}

\caption{Plots of the dispersion relations for (a) Aluminum and (b) Chromium solid films with thicknesses
of $1\, nm$ and $2\, nm$, using and estimated Hamaker coefficient $A=5\times10^{-19}\, J$.
Both types of films are stable to small perturbations of all wavelengths for $h_{o}\ge\sim2\, nm$.
\label{fig:AlandCrDispersionPlots}}
\end{figure}

In the case of a $1\, nm$ film, the stability of the solid film is strongly dependent on the
choice of the Hamaker coefficient $A$, which dictates the magnitude of the disjoining pressure.
When using the experimentally determined value of $A=1.4\times10^{-18}\, J$ \cite{favazza06d}
in eq. \ref{eq:sigmaSolidState}, the film is unstable and there exists a characteristic wavelength
that will grow the fastest. The dispersion curve for growth/decay rates of perturbations in the
solid state has been shifted to larger wavelengths, as compared to that for the TFH instability,
as shown in Figure \ref{fig:CoDispersionPlots}(a). In this case the cut-off  and the fastest
growing wavelengths have been increased due to the stabilizing effects of volume diffusion and
evaporation-condensation. In the absence of of these effects $k_{max}$ and $\lambda^{*}$ are
the same for both the solid and liquid states. The importance of the magnitude of the dispersion
term on mass transport can be seen if one uses the lower limit of the Hamaker coefficient for
metals $(\sim10^{-19}\, J)$ \cite{Isaelachvili}. When a theoretically estimated $A\sim3\times10^{-19}\, J$
was employed, the disjoining pressure was not large enough to destabilize the solid film; but
rather, the film was seen to be stable against perturbations of all wavelengths. For a $10\, nm$
film, the dispersion forces with $A=1.4\times10^{-18}\, J$ are not large enough to instigate
a dewetting instability in the solid film, as shown if Figure \ref{fig:CoDispersionPlots}(b).
For the length scales stemming from a TFH instability, $\mid\sigma_{L}\mid\gg\mid\sigma_{S}\mid$,
thereby confirming that hydrodynamic flow is the more dominant mass transport mechanism during
$ns$ pulsed laser melting%
{}. We obtained similar results for a variety of metals suggesting that this is a general result,
applicable to most thin metal films. Most metals investigated have similar diffusion rates, but
the evaporation rates vary greatly. Figs. \ref{fig:AlandCrDispersionPlots}(a-b) show plots of
dispersion curves for two extreme scenarios, one metal (Al) with a low evaporation rate and the
other (Cr) with a large rate. In both cases, films with $h_{o}\ge2\, nm$ are seen to be stable
to small perturbation of all wavelengths. 

There are several implications manifested in these results. First, the timescales of solid state
process are much larger than the timescales associated with hydrodynamic flow. Therefore, the
\emph{quenched-in} structures from the pulsed laser melting and resolidification are likely unchanged
during the cooling period. This conclusion is corroborated by experimental observations of pattern
formation in ultrathin metal films by using multiple pulses of  ($266\, nm$ wavelength, $9\, ns$
pulsed) ultraviolet (UV) laser irradiation: see \cite{Favazza06b,favazza06c,favazza06d,Trice06a,Favazza07a}
for details. A variety of metal thin films, including Co, Fe, Cu, Ag, Ni and Pt, were deposited
under ultrahigh  vacuum ($\sim10^{-8}$ Torr) by e-beam evaporation or pulsed laser deposition
onto commercially obtained SiO$_{\text{2}}$/Si substrates with the thermally grown oxide layer
of 400 nm thickness. Film thicknesses ranged from $1-10$ nm. The initial films had an average
surface roughness $\leq0.2\, nm$ as evaluated by atomic force microscopy. Following deposition,
the films were irradiated at various laser energies above a threshold energy density $E_{th}$
such that appreciable surface deformation was observed for various irradiation times, as measured
by the number of pulses $n$ ranging from $10\leq n\leq10,500$. Following such irradiation,
the surface morphology and concentration of metal was evaluated by imaging and energy dispersive
X-ray spectrometry in a scanning electron microscope, respectively. It was found that all the
metals possessed a thickness-dependent threshold energy density $E_{th}$ above which substantial
surface changes were observed. However, for energies below this threshold, no surface roughening
was visible even after the longest irradiation times of 10,500 pulses. This energy threshold
has been attributed to the energy required to melt the films \cite{Trice06a}.  By irradiating
films with a laser energy density above the melt threshold, we performed detailed studies of
the dynamics of the pattern evolution for Co films as a function $n$ and found that the dewetting
patterns showed characteristic film-thickness dependent length scales and spatial order consistent
with the predictions of eq. \ref{eq:sigmaLiquid}. Hence, despite the much larger time spent
in the solid phase (i.e. cooling period), the overall contribution from solid state transport
was inferred to be negligible.

\section{Thermal Stress\label{sec:Thermal-Stress}}

Since we focus on the dynamics of perturbed thin films on amorphous substrates, the prevalent
stresses that are present in the film are thermally generated. Due to the mismatch in the thermal
expansion coefficients between the film and the substrate, the film is under compressive stress
while it is heated and a tensile stress when it is cooling.  This stress is given by the following
expression: \begin{equation}
\sigma_{f}=\frac{(\alpha_{s}-\alpha_{f})\Delta T}{h_{f}[\frac{1-\nu_{f}}{h_{f}E_{f}}+\frac{1-\nu_{s}}{h_{s}E_{s}}]}\label{eq:ThermalStress}\end{equation}
where $\alpha$ is the thermal expansion coefficient, $\nu$ is Poisson's ratio, $E$ is Young's
modulus, $h$ is the thickness of the layer, $\Delta T$ is the temperature change, and the subscripts
$f$ and $s$ denote the film and substrate quantities, respectively \cite{Ohring}. In the limiting
case where $\frac{h_{s}E_{s}}{1-\nu_{s}}\gg\frac{h_{f}E_{f}}{1-\nu_{f}}$, the preceding expression
reduces to: \begin{equation}
\sigma_{f}=\frac{(\alpha_{s}-\alpha_{f})}{1-\nu_{f}}\Delta T.\label{eq:ThermalStressReduced}\end{equation}
Stresses related to thermal expansion for most thin metal films on a silica substrate are on
the order of a few $GPa$. However, they are still lower than the critical value required to
initiate a buckling instability in the film, which is given by the following equation:\begin{equation}
\sigma_{critical}=\phi(\frac{E_{f}}{1-\nu_{f}^{2}})^{\frac{1}{3}}(\frac{E_{s}}{1-\nu_{s}^{2}})^{\frac{2}{3}}\label{eq:crticial stress}\end{equation}
%
{}where $\phi=0.52$ is a constant. Typical critical stress values range from $\sim40-60\, GPa$.
Stress can also result from an anisotropic temperature distribution in the film \cite{Dawson}.
This could arise in pulsed laser processing. As described above, the thickness variations caused
by the TFH instability are quenched in during the rapid resolidification phase. Hence, in the
solidified deformed film, a non-uniform temperature profile could prevail for times comparable
to the diffusion time scale \cite{Trice06a}. Within the context of linear theory of elasticity,
the thermoelastic stress will also decay at the same rate as the the temperature profile in the
film \cite{Dawson}. Hence a characteristic time scale of temperature/stress diffusion can be
estimated by solving the transient heat equation for a 2-dimensional film:\begin{equation}
\rho_{f}C_{P,f}\frac{\partial T}{\partial t}=\kappa_{f}\frac{\partial^{2}T}{\partial x^{2}}+\kappa_{f}\frac{\partial^{2}T}{\partial z^{2}}\label{eq:HeatEq}\end{equation}
where $\rho_{f}$, $C_{P,f}$, and $\kappa_{f}$ are the density, heat capacity and thermal conductivity
of the film, respectively. Assuming an initial sinusoidal temperature profile in the lateral
(\emph{x}) direction and a uniform temperature distribution throughout the thickness (\emph{z}
direction) of the film, one finds the following relationship for the time dependent component
of the temperature, $T'(t)$, in which the temperature, $T(x,t)=T_{0}(x)T'(t)$:\begin{equation}
T'(t)=exp[-\alpha_{f}k^{2}t-\frac{2c\sqrt{\rho_{s}C_{P,s}\kappa_{s}}}{\rho_{f}C_{P,f}h_{0}}\sqrt{t}]\label{eq:TempProfile}\end{equation}
$\alpha_{f}$ is the thermal diffusivity of the film, $k$ is the wavevector of the perturbation,
$c$ is a constant between $0$ and $1$ and $\rho_{s}$, $C_{P,s}$, and $\kappa_{s}$ are the
density, heat capacity and thermal conductivity of the substrate, respectively. Since the length
scale of the periodic thickness variations is being imprinted on the film in the liquid state,
one can take the expected value of $k$ for a corresponding $h_{o}$ from TFH theory. Using these
values $(c=0.2)$, the timescale during which this stress could influence solid state mass transport
is on the order of a picosecond for a perturbation wavelength, $\lambda\sim30\, nm$ and on the
order of a nanosecond for $\lambda\sim1\,\mu m$. These timescales are too short to appreciably
effect any solid state process in modifying the surface of the film.

\section{Conclusion}

We have developed a mathematical model to describe thin film dynamics on amorphous surfaces by
incorporating the effects of surface/volume diffusion, evaporation-condensation and dispersion
forces. New results are presented for volume diffusion in thin films. Specifically, for $kh_{0}$<\,{}<
1, the decay rate of surface perturbations by this mechanisms is proportional to $k^{2}/h_{0}$
as compared to $k^{3}$ for bulk materials. It is shown that depending on the magnitude of the
Hamaker coefficient, ultrathin solid films with thickness < 2 nm could be prone to pattern forming
instabilities while films with thickness > 2 nm are predicted to be stable for a variety of metals.
 The model was applied to analyze the role of solid state mass transport mechanisms on the dynamics
of thin films processed by pulsed laser irradiation. The upper limits for the time constants
associated with the decay of surface perturbations in the solid state was found to be much smaller
compared to the time interval between subsequent pulses. Further, a qualitative analysis of thermoelastic
stress dissipation in the film-substrate bilayer system showed that the time scale of stress
relaxation is much smaller than the pulse duration. These findings suggest that contributions
to the changes in surface deformation from solid state processes are minimal and the pattern
formation occurs through a hydrodynamic instability that is followed by laser melting of the
film, an observation corroborated by experimental observations.

\section*{Acknowledgements}

RK and RS acknowledges support by the National Science Foundation through grant CAREER DMI-0449258
and CTS-0335348, respectively. CF acknowledges valuable contributions from Hare Krishna.


\pagebreak

\end{document}